\documentclass{article}
\usepackage{spconf,amsmath,amsfonts, graphicx,multirow}
\usepackage{subcaption}
\usepackage{epstopdf}

\title{Reformulating Speaker Diarization as Community Detection With Emphasis On Topological Structure}
\name{Siqi Zheng, Hongbin Suo}
\address{Speech Lab, Alibaba Group\\
\small\texttt{\{zsq174630, gaia.shb\}@alibaba-inc.com}}

\begin{document}
\ninept
\maketitle
\begin{abstract}

Clustering-based speaker diarization has stood firm as one of the major approaches in reality, despite recent development in end-to-end diarization. However, clustering methods have not been explored extensively for speaker diarization. Commonly-used methods such as k-means, spectral clustering, and agglomerative hierarchical clustering only take into account properties such as proximity and relative densities. In this paper we propose to view clustering-based diarization as a community detection problem. By doing so the topological structure is considered. This work has four major contributions. First it is shown that Leiden community detection algorithm significantly outperforms the previous methods on the clustering of speaker-segments. Second, we propose to use uniform manifold approximation to reduce dimension while retaining global and local topological structure. Third, a masked filtering approach is introduced to extract ``clean" speaker embeddings. Finally, the community structure is applied to an end-to-end post-processing network to obtain diarization results. The final system presents a relative DER reduction of up to 70 percent. The breakdown contribution of each component is analyzed. 

\end{abstract}
\section{Introduction}

End-to-end speaker diarization has attracted heated discussion recently~\cite{DBLP:conf/asru/FujitaKHXNW19}\cite{DBLP:conf/interspeech/FujitaKHNW19}. Despite its strenghts in handling overlapping speech, its limitations in processing long meetings and large number of speakers have hindered its large-scale usage. Considering these limitations, some researchers have spent considerable efforts in exploring possibilities to incorporate clustering-based approaches with end-to-end diarization in order to take advantage of both systems ~\cite{DBLP:conf/icassp/HoriguchiGFWN21}\cite{DBLP:conf/icassp/KinoshitaDT21}\cite{DBLP:journals/corr/abs-2105-09040}. 

Clustering-based approaches are known to be more robust on large meetings and cross-domain dataset~\cite{DBLP:conf/asru/FujitaKHXNW19}\cite{DBLP:journals/corr/abs-2105-09040}. Hence these methods are so far irreplaceable in real-world applications of speaker diarization. However, most diarization systems seem to have decided to save the  efforts of finding the optimally performing clustering methods and readily accepted some of the most common practices such as k-means, spectral clustering, and agglomerative hierarchical clustering (AHC)~\cite{DBLP:conf/interspeech/McCreeSG19}\cite{DBLP:journals/spl/ParkHKN20}\cite{DBLP:conf/interspeech/LinYLBB19}.

Despite their simplicity and popularity, these clustering methods may not be most suitable for the sake of clustering speaker-segments. For example, k-means only considers the relative distance to the centroids of clusters and fails to recognize the topological structure of the distribution. Spectral clustering, like k-means, have difficulties determining the number of classes, which is essential in a speaker diarization system.

In this work we propose to reformulate the clustering of speaker-segments as a network community detection problem. By traversing the network we seek to understand its entire structure. When it comes to the decision making for each node, both local and global topology are considered. The movement of a node to another community will have impact on community structure of the whole network. The objective is to optimize such impact.

The Louvain algorithm has become one of the mainstream community detection algorithms since its proposal~\cite{Blondel_2008}. Its popularity rises rapidly in various areas of studies due to its simplicity and effectiveness~\cite{DBLP:journals/corr/abs-0906-0612}\cite{DBLP:journals/neuroimage/RubinovS10}. However, it is proven that the Louvain algorithm could result in arbitrarily badly connected communities. In terms of clustering of speaker-segments, this would lead to false aggregation of two speakers into one. In this paper we introduce the recently proposed Leiden algorithm, which guarantees that all communities are well-connected~\cite{DBLP:journals/corr/abs-1810-08473}. 

Dimension reduction also has a positive effect on clustering of speaker embeddings. A speaker embedding encodes information such as voice, speech content, channel, environment, etc \cite{DBLP:conf/interspeech/ZhengS20}. The embedding extraction network is trained on the set of hundreds of thousands of speakers. A high-dimensional embedding space allows for larger margins between large number of distinct classes during training. During inference, however, we only need to distinguish among a few speakers presented in a meeting. Hence information encoded in an embedding may be redundant or sometimes harmful due to curse of dimensionality. It is of our interest to project embeddings onto lower dimension space that best serves to distinguish between the speakers in the given meeting. In this paper we propose to use a novel dimension reduction technique named Uniform Manifold Approximation and Projection (UMAP)~\cite{DBLP:journals/jossw/McInnesHSG18}. This manifold learning technique stems from the theoretical foundations in Riemannian geometry and algebraic topology and has proven to be effective in our task. 

Segmentation is a long-standing challenge for clustering-based diarization~\cite{DBLP:conf/icassp/Garcia-RomeroSS17}\cite{DBLP:conf/icassp/WangDWMM18}\cite{DBLP:conf/icassp/ZhengHWSFY21}. Longer segments may contain multiple speakers, which could generate noisy embeddings for clustering. On the other hand, shorter segments may result in large variance in embeddings, hence introducing noise to clustering. To deal with this dilemma, a masked filtering technique is presented. The filtering process adopts a ``winner takes all" strategy. The dominating speaker is chosen as the target speaker for the segment and the corresponding embedding is used as target reference for the extraction of clean embedding.

Upon determining the community partitions for all speaker-segments, we seek to further improve the performance by refining the results within and between consecutive segments, such as locating precise speaker change points, smoothing out frame-level results, and handling overlapping segments. These tasks can be addressed by an end-to-end post processing component. In this work we extend the end-to-end mechanism for post processing that effectively integrates the partition results from community structure.

\label{sec:intro}

\section{Methods}

\subsection{Leiden Community Detection}

The construction of a community detection network is straightforward. Speaker embeddings extracted from each segments are formulated as nodes in the network and similarity scores between pairs of embeddings are the edge values. The goal is to find the optimal partitions that describe the community structure. 

Modularity is selected as the optimization function and is represented as 
$$\mathcal{Q}=\frac{1}{2m}\sum_i\sum_j(A_{ij}-\frac{k_ik_j}{2m})\delta(C^{(i)},C^{(j)}),$$

where A is the adjacency matrix, $m = |E|$ is the number of edges in the total network, $k_i$ is the degree of vertex $i$, $C^{(i)}$ is the community vertex $i$ belongs, and $\delta(C^{(i)},C^{(j)})$ is the Kronecker delta function that equals 1 if $C^{(i)}=C^{(j)}$ and 0 otherwise.

The Leiden algorithm works as follows:

\textbf{Step 1.} For each node whose neighborhood has changed, move to a different community until no movement of nodes can increase modularity. This results in a partition $\mathcal{P}$.

\textbf{Step 2.} Make refinements to $\mathcal{P}$: Within each community in $\mathcal{P}$, set to singleton partition and locally merge nodes to form $\mathcal{P}_{refined}$.

\textbf{Step 3.} Create aggregated network based on the refined partition.

\textbf{Step 4.} Repeat Step 1-3 until no further improvements can be made.

\subsection{Uniform Manifold Approximation}

UMAP presumes the existence of a locally connected manifold on which the speaker embeddings are uniformly distributed and aims to preserve the topological structure of this manifold. 

Similar to other dimension reduction algorithms, the UMAP mechanism can be summarized by two stages. First it seeks to construct a suitable weighted k-neighbour graph. Then it estimates the projection of low dimensional space of the graph. 

Let $X=\{x_1,\ldots,x_N\}$ be the embeddings of speaker-segments in the meeting of interest. The weights between an embedding to its k nearest neighbours are given by 

$$w((x_i,x_{i_j}))=\exp\Big(\frac{-max(0,d(x_i,x_{i_j})-\rho_i)}{\sigma_i}\Big)$$

\noindent for $1\leq i \leq N, 1\leq j \leq k$. Here $\rho_i$ exemplifies local-connectivity assumption and $\sigma_i$ is a normalization factor defining the Riemannian metric on which $x_i$ lies on such that the data is uniformally distributed. 

A weighted graph can thus be constructed based on the above weight function. Theoretically the graph is the 1-skeleton of the fuzzy simplicial set related to the Riemannian metric space ambient to $x_i$. Let $A$ be the adjacency matrix of this graph.

UMAP is optimized through the fuzzy set cross entropy given by 

$$C = \sum_{a\in A}\mu(a)\log\Big(\frac{\mu(a)}{\upsilon(a)}\Big)+(1-\mu(a))\log\Big(\frac{1-\mu(a)}{1-\upsilon(a)}\Big),$$

\noindent where $\mu$ and $\upsilon$ are the membership functions of two fuzzy sets. Note that $C$ is written as the sum of two components. The first component on the left of summation serves to find local clustering, or high density regions in the graph. The second component aims to preserve the global topological structure. 

The t-SNE algorithm had been one of the most popular dimension reduction algorithms before UMAP was proposed~\cite{vanDerMaaten2008}.  However, t-SNE fails to preserve the  structure like UMAP does. In addition, t-SNE only allows for reduction to 2-dimensional space while UMAP allows for reduction to any dimension that is optimal for the data of interest. Therefore, UMAP is more suitable for tasks requiring further processing, such as clustering, after dimension reduction. 

\subsection{``Winner takes all" Masked Filtering}

We also seek to make improvements to the basic unit of community detection -- the speaker embeddings. Most clustering-based speaker diarization systems split the audio into segments of length ranging from 1 to 2 seconds. This is based on the presumption that a short 2-second segments will contain no more than 1 speaker. As the length of the segments increase, such presumption no longer holds. In our system we audaciously increase the length of each segments to 4 seconds, with the understanding that longer segments lead to more robust embeddings. 

\begin{figure}[h!] 
\centering
  \includegraphics[width=\linewidth]{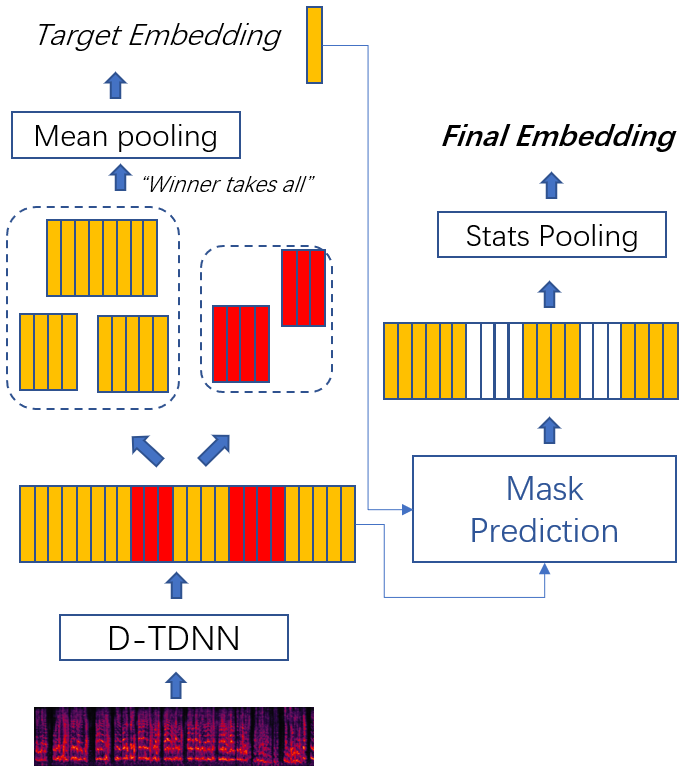}
  \caption{An illustration of the ``Winner-takes-all" masked filtering system. }
  \label{CAM}
\end{figure}

In order to extract relatively clean embeddings from longer segments, we introduce a ``winner-takes-all" masked filtering system as illustrated in Figure \ref{CAM}. The filterbank features of each segment is passed through a several D-TDNN layers. Then the frame-level speaker features are clustered into 2 classes using k-means clustering. Features from the dominating component are used as the target embedding. The masked prediction network is trained in the same manner as described in \cite{DBLP:conf/icassp/YuZSLL21}. 

One may wonder why bother doing masked filtering and not just use the \textit{Target Embedding} in Figure \ref{CAM} as the output. The mask prediction component serves two purposes. First, frames from the winner speaker may still contain overlapping speech from other speakers, which will introduce bias to the target embedding. Masked filtering allows us to extract relatively clean embeddings from overlapping speech. This is well illustrated in our previous work\cite{DBLP:conf/icassp/YuZSLL21}. Second, if there is only one speaker in the entire segment, it is of our best interest to include as much information as possible. By design the target embedding could ignore up to half of the frames with useful information. This contradicts our motivation of extracting embeddings from longer segments in order to minimize variance. 

\subsection{End-to-end Post Processing with Community Partitions}

Despite that clustering-based diarization has superior performance in determining the number of speakers, finding correct classification for speaker-segments, and handling domain mismatch between training and testing data, we recognize that end-to-end approaches are more suitable at processing local information such as handling overlapping speech, and finding precise speaker change points. Therefore we propose to include an end-to-end post processing component upon the results of community detection. 

The end-to-end post processing architecture is a modification from the EEND-EDA system described in \cite{DBLP:conf/interspeech/HoriguchiF0XN20}, in which the Encoder-Decoder Attractor(EDA) component stores a flexible number of speakers. As indicated by the experiments, the EDA component is limited in predicting correct number of speakers. Hence we replace the EDA module by the results of community partitions and fix the number of speakers as the number of communities. During training and inference, the zero vectors input of EDA are replaced by the mean of each community parititons and the generation of new attractors is disallowed. For the simplicity of referencing, we name this end-to-end post processing approach based on community partitions as EEND-CommPart in later sections. 

Since EEND-CommPart is aimed for processing local information, the input to EEND-CommPart are short segments. In our experiment, for example, the inputs consist of two adjacent segments, making the total length 8 seconds. During training longer segments are included to increase variety.

\section{Experiments \& Results}
\subsection{Corpus}

The train and test set used in the experiments are simulated from NIST SRE corpus. The training corpus consists of 57,517 utterances from 5,767 speakers in NIST SRE 04-10 corpus. The performance is evaluated on SRE10 Evaluation set.  When evaluating diarization performance, utterances from a random 2-8 speakers are selected to simulate meetings of duration of at least 30 minutes. The overlap ratio ranges from $5\%$ up to $30\%$ in each of the  simulated meetings.

\subsection{Experiments}

In the first experiment we evaluate performance of several diarization systems on simulated long meetings with 2, 4, 6, and 8 participants respectively. The components described above, such as ``winner takes all" masked filtering, UMAP-Leiden community detection, and EEND-CommPart, are included or excluded in different runs in order to see their effects on the overall performance.

In the second experiment we focus on evaluating the performance of several clustering approaches. Speaker-segments ranging from 2-4 seconds are randomly selected. 

Specifically we compare k-means, spectral clustering, AHC, and UMAP-Leiden. We are also interested to see how EEND-EDA performs on predicting the number of speakers compared to other clustering approaches.

For k-means and spectral clustering, we use the curve of descending eigenvalues to estimate the number of classes, as described in \cite{DBLP:conf/nips/NgJW01}. The number of classes of AHC is counted once the stopping criteria is met. The stopping criteria is determined on a small development set. Three hyperparameters are required to be fine-tuned for UMAP dimension reduction algorithm - the number of k nearest neighbors to be considered, the target dimension to be reduced to, and expected distance that determines whether points should be packed together. For Leiden algorithm the most important hyperparameter to be fine-tuned is the resolution. 

The mask prediction network is trained on the same setup as described in \cite{DBLP:conf/icassp/YuZSLL21}. The only difference is that, instead of taking a rough mean pooling as the target embedding, we conduct a 2-class clustering and only run mean pooling on the dominating class. 

The speakers embeddings are extracted from a D-TDNN architecture~\cite{DBLP:conf/interspeech/YuL20} trained on the NIST SRE 04-10 train set~\cite{nistsre10}. AM-Softmax is used as the loss function~\cite{DBLP:conf/icassp/YuFL19}.

\subsection{Results \& Analysis}

As Table \ref{tab:Result1} suggests, system No.6 has the optimal performance when there are 4 or more speakers. System No.6 includes masked filtering as pre-processing, UMAP-Leiden as clustering, and EEND-CommPart as post-processing. System No.2, EEND-EDA, has optimal performance for 2-speaker cases, but its performance decreases significantly as the number of speakers increases. System No.6 outperforms EEND-EDA and k-means clustering baseline by a remarkable margin in 4-, 6-, and 8-speaker cases.

When compared the performance of system No.1 to No.4, as well as system No.3 to No.5, we can see that UMAP-Leiden introduces remarkable gains in performance. System No.5 adds masked filtering onto system No.4, showing that obtaining cleaner embeddings has positive effects. Finally, the results from system No.6 indicate that EEND-CommPart post processing allows for better handling of overlapping speech and better local refinements.

\begin{table*}[tb]
    \caption{Performance comparison of various diarization systems on simulated long meetings from NIST SRE corpus.}
    \centering
    
    \begin{tabular}{cccccccc}
    \hline
      \multirow{2}{*}{System ID}  & \multirow{2}{*}{Pre-Processing} &  \multirow{2}{*}{Clustering Methods} & \multirow{2}{*}{Post-Processing} & \multicolumn{4}{c}{DER(\%)} \\
       & & & & 2spk & 4spk & 6spk & 8spk \\
      \hline
       No.1  & /  & k-means & / & 10.3  & 23.9 & 33.6 & 40.1\\ \hline
    No.2 & /  & EEND-EDA & / & \textbf{4.7}  & 22.8 & 42.3 & 68.5 \\ \hline
    No 3 & Masked Filtering & k-means & / & 9.6  & 23.0 & 33.8 & 37.2\\ \hline
    No.4 & / & UMAP-Leiden & / & 7.3   & 17.1 & 24.6 & 28.9 \\ \hline
    No.5& Masked Filtering & UMAP-Leiden & / & 6.5 & 15.9  & 22.3 & 25.8 \\ \hline
    No.6 & Masked Filtering & UMAP-Leiden & EEND-CommPart & 5.2 &  \textbf{13.1} & \textbf{18.8} & \textbf{20.2}\\ \hline
    \end{tabular}
    \label{tab:Result1}
\end{table*}

Now we break down to analyze the gain from clustering performance of UMAP-Leiden algorithm. 

Table \ref{tab:cluster} displays the performance of different clustering methods when the actual number of speakers is 1, 2, 4, 6, 8, and 10, respectively. For each case 500 clustering tests are simulated. For example, for \#Spks=2, random shuffled segments from a 2 random speakers are selected to construct a clustering test instance. This is repeated 500 times. The F-score and \#Spks Prediction Accuracy are estimated based on the average of 500 clustering results. \#Spks Prediction Accuracy measures the number of times out of total that the predicted number of speakers equals the actual number of speakers. 

According to Table \ref{tab:cluster}, when the actual number of speakers in the meeting is either 1 or 2, k-means, spectral clustering, and UMAP-Leiden have similar competitive performances, with spectral clustering slightly better for 1-speaker case and k-means slightly better for 2-speaker case. The competitive performance of k-means and spectral clustering for diarization on very few speakers may be a major reason that researchers lack the motivation to explore more sophisticated clustering methods for speaker diarization.

As the number of speakers increase, the performance of k-means and spectral clustering drop significantly, largely due to the difficulties in estimating correct number of speakers. The UMAP-Leiden algorithm, on the other hand, performs reasonably well on larger number of speakers. We also note that EEND-EDA performs well for 2-speaker situation and has relatively poor performance on estimating the number of speakers for \#spks $> 2$. This is consistent with the observation from Table \ref{tab:Result1} that EEND-EDA has the lowest DER for \#spk=2 but increases rapidly. This is our main motivation for replacing the EDA module by the results of UMAP-Leiden partitions in our end-to-end post processing component.

\begin{table}[htbp]
  \centering
  \caption{Comparison of speaker clustering performance for various number of speakers.}
    \begin{tabular}{clcc}
    \hline
    \#Spks & Methods & \#Spk Predict Accuracy & F-score \\
    \hline
    \hline
    
    \multirow{4}[1]{*}{1} & K-means & 0.84  & 0.91    \\
          & \textbf{Spectral} & \textbf{0.84}  & \textbf{0.93}    \\
          & AHC   & 0.32  & 0.64    \\
          
          & UMAP-Leiden   & 0.71    & 0.87  \\
    \hline
    \multirow{4}[2]{*}{2} & \textbf{K-means} & \textbf{0.95}  & \textbf{0.95}   \\
          & Spectral & 0.95  & 0.94   \\
          & AHC   & 0.77  & 0.86    \\
          & EEND-EDA   & 0.93  & NA    \\
          & UMAP-Leiden   & 0.93   & 0.92  \\
    \hline
    \multirow{4}[2]{*}{4} & K-means & 0.83  & 0.86   \\
          & Spectral & 0.83  & 0.88  \\
          & AHC   & 0.70  & 0.85   \\
          & EEND-EDA   & 0.50  & NA    \\
          & \textbf{UMAP-Leiden}   & \textbf{0.90}   & \textbf{0.94}  \\

    \hline
        \multirow{4}[2]{*}{6} & K-means & 0.62  & 0.81  \\
          & Spectral & 0.62   & 0.84  \\
          & AHC   & 0.61  & 0.84    \\
          & EEND-EDA   & 0.31  & NA    \\
          & \textbf{UMAP-Leiden}   & \textbf{0.85}   & \textbf{0.89}  \\

    \hline
        \multirow{4}[2]{*}{8} & K-means & 0.45  & 0.70   \\
          & Spectral & 0.45   & 0.72  \\
          & AHC   & 0.55  & 0.79   \\
          & EEND-EDA   & 0.15  & NA    \\
          & \textbf{UMAP-Leiden}   &
          \textbf{0.84}  & \textbf{0.87}  \\

    \hline
        \multirow{4}[2]{*}{10} & K-means & 0.27  & 0.67  \\
          & Spectral & 0.27   & 0.69  \\
          & AHC   & 0.46  & 0.74   \\
          & EEND-EDA   & 0.03  & NA    \\
          & \textbf{UMAP-Leiden}   &
          \textbf{0.80}  & \textbf{0.84}  \\

    \hline
    \end{tabular}%
  \label{tab:cluster}%
\end{table}%

Table \ref{tab:runtime} compares the computation costs of different clustering approaches. The community detection approaches such as Louvain and Leiden are most efficient in terms of running time. Leiden is faster than Louvain because it adopts a fast local moving process which visits only nodes whose neighbourhood has changed, while Louvain keeps visiting all nodes in the network every single time.

\begin{table}[h]
    \caption{Comparison of computations costs of various clustering methods, evaluated on 150,000 data points.}
    \centering
    \begin{tabular}{|c|c|}
    \hline
      Methods   & Runtime \\
      \hline
    k-means & 2585s \\ \hline
    Spectral  & 597s \\ \hline
    AHC  & 3281s \\ \hline
    Louvain  & 374s \\ \hline
    \textbf{Leiden}  &\textbf{171s} \\
    \hline
    \end{tabular}
    \label{tab:runtime}
\end{table}

\section{Conclusion}

In this paper we propose to reformulate speaker diarization as community detection problem. We introduce Leiden as the community detection algorithm for its optimal performance on the dataset used in this experiment. Other community detection algorithms, such as Infomap~\cite{Rosvall_2009} and Louvain, have demonstrated to be slightly better on some datasets. Therefore, deciding on which algorithm to use is dependent on the data of interest. Fortunately, implementations of all above mentioned community detection and dimension reduction algorithms are relatively simple. A quick comparison can be done with minimal efforts.

We find ways to extract speaker embeddings from longer segments to reduce variance, without having affected by other speakers presented in the segment, by introducing a masked filtering approach. This turns out to be more beneficial in meetings presented with large portions of overlapping speech and frequent speaker turns. Considering its computation costs, the masked filtering can be optionally turned off for more structured and organized meetings. For the future we are interested in exploring more efficient methods to serve the purpose. 

Finally, we propose the EEND-Commpart post processing component to handle overlapping speech and polish local results. The system leverages the edge of end-to-end methods in detailed refinement and the advantage of clustering-based approaches in global speaker counting.

Through this work we show the clustering-based diarization approaches still have large rooms for improvement. We observed remarkable gains simply by replacing the k-means with community detection algorithms. We hope that this result could inspire more studies on the clustering methods for speaker diarization. 

\bibliographystyle{IEEEbib}
\bibliography{refs,strings}

\end{document}